\documentclass[apl,twocolumn,showpacs]{revtex4-1}

\usepackage{graphicx}% Include figure files
\usepackage{dcolumn}% Align table columns on decimal point
\usepackage{bm}% bold math
\usepackage{natbib}
\usepackage{amsmath} 
\usepackage{epstopdf}
\usepackage{epsfig}
\usepackage{lipsum}

\usepackage{amssymb}  % for math spacing
\usepackage{amsfonts}
\usepackage{url}  % Hyphenation of URLs.
\usepackage[justification=raggedright]{caption}	% makes captions ragged right - thanks to ryce Lobdell

\usepackage{fixltx2e}  % subscript
\usepackage{paralist}  % for compactitem and compactenum
\usepackage{siunitx}   % for scientific notation

\usepackage[labelfont=bf]{caption} % Bold figure number
\usepackage{textgreek} % Non-italic Greek letters

\newcommand{\um}{{\textmu}m }

\usepackage{url}
\usepackage{placeins}
\usepackage[usenames,dvipsnames]{color}    % ADDED FOR COLOR NAMES
\usepackage{soul}   % SOUL Added for underlines
\setstcolor{red} % strikethrough color
\setul{}{1.5pt}  % line thickness

\draft % marks overfull lines with a black rule on the right

\begin{document}

%\preprint{}

% \title{Viscosity Sensing Using Optomechanics} %Title of paper
\title{Optomechanical non-contact measurement of microparticle compressibility in liquids}

\author{Kewen Han}
\email{hankewen1@gmail.com}
\author{Jeewon Suh}
\author{Gaurav Bahl}
\email{bahl@illinois.edu}
\affiliation{Mechanical Science and Engineering, University of Illinois Urbana Champaign, 1206 W. Green St., Urbana, IL 61801, USA}

\date{\today}

\begin{abstract}
High-throughput label-free measurements of the optical and mechanical properties of single microparticles play an important role in biological research, drug development, and related large population assays. 
Mechanical detection techniques that rely on the density contrast of a particle with respect to its environment are blind to neutrally bouyant particles.
However, neutrally buoyant particles may still have a high compressibility contrast with respect to their environment, opening a window to detection.
Here we present a label-free high-throughput approach for measuring the compressibility (bulk modulus) of freely flowing microparticles by means of resonant measurements in an opto-mechano-fluidic resonator.

\end{abstract}

\pacs{42.60.Da, 42.81.Pa, 07.07.Df}% insert suggested PACS numbers in braces on next line

\maketitle %\maketitle must follow title, authors, abstract and \pacs

The mechanical properties of individual cells (e.g. density, Young's modulus, and compressibility) are important in determining how they interact with their environment.  Variations of mechanical properties are also known to correlate with specific disease states including anemia \cite{Mohandas1994a}, malaria \cite{Suresh2005}, and cancers \cite{Suresh2005}, and can influence cell differentiation \cite{Gonzalez-Cruz2012}.
While optical flow cytometry is an extremely powerful tool for single cell analysis and sorting \cite{Hedley1983, Cho2010}, an equivalent high-throughput single cell mechanical assay is not yet available. Knowledge of mechanical parameters  of single microparticles or  cells could enable new discoveries and aid the development of next generation diagnostics.

Traditionally, measurement of mechanical properties of individual microparticles requires the application of forces, upon which  the mechanical responses (e.g. deformation) can be quantified to determine the properties of interest. 
Such forces can be applied through direct contact techniques like AFM deformational probing \cite{Cross2007,Kuznetsova2007b}, optical tweezing of adhered beads \cite{Lim2004}, micropipette aspiration \cite{Hochmuth2000}, and mechanical resonator loading \cite{Malvar2016,Corbin2016,Park2015}.  
These methods, however, are inherently slow since the analyte particle must be temporarily immobilized.
In contrast, flow-through type sensors can offer much higher throughput measurements, 
 for instance, single cell mass can be rapidly measured by flowing it through an internal channel within a mechanical resonator \cite{Olcum2015}. 
Such  methods, however, have so far relied on the density contrast of a particle with its environment and therefore cannot detect neutrally buoyant particles. The solution to detecting neutrally buoyant particles lies in their compressibility contrast against their environment, which can open a window to detection. 
Recently, Hartono et. al. demonstrated \cite{Hartono2011} a non-contact compressibility measurement  {for single cells} using the acoustic radiation force and cell trajectories.
However, the direct measurement of single microparticle compressibility with high throughput has not yet been achieved.

\begin{figure}[t]
	\centering
	\includegraphics[width=0.45\textwidth, clip=true, trim=0in 8.65in 5in 0in]{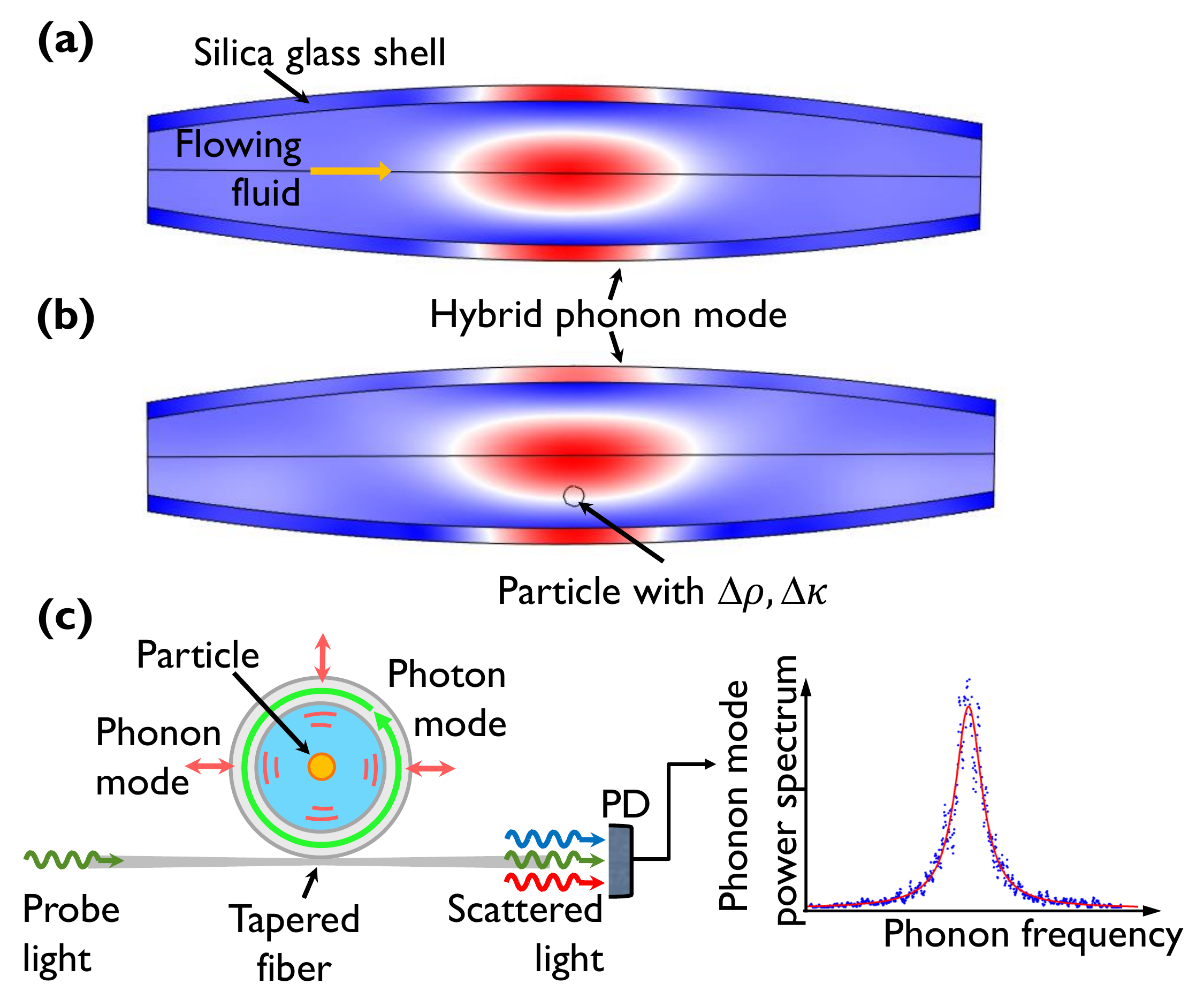}
	\caption{{\bf  } 
		(a) A fluid-shell hybrid breathing mechanical (phonon) mode in an opto-mechano-fluidic resonator (OMFR). 
		(b) Particles of density contrast $\Delta \rho$ and compressibility contrast $\Delta \kappa$ change the vibrational mode shape as indicated by the slight broken symmetry.  
		(c) The thermal mechanical fluctuations of the OMFR mode can be measured optically \cite{Han2016} via a single-point tapered fiber measurement. Analysis of this spectrum conveys information on the particle.
	 } 
	\label{fig:Principle}
\end{figure}

In this work, we report a new acoustic-based high-throughput technique for measuring the compressibility of single particles without physical contact, using an opto-mechano-fluidic resonator (OMFR) \cite{Bahl2013a,HyunKim2013}. 
These fused silica microcapillary resonators are cavity optomechanical sensors \cite{Gil-Santos2015,Kim2017,Wu2016,Yu2016,Kaminski2016} that support ultrahigh-Q optical modes coupled to co-localized mechanical (phonon) modes of the structure. Fluid analytes can be flowed internally without influencing the optics (Fig. \ref{fig:Principle}(a)). Phonons within the mechanical resonant mode permeate the entire cross-section of the capillary, including the fluid, casting a near-perfect net for measuring particles flowing inside.
All particles in the sample must transit and perturb the phonon modes (Fig. \ref{fig:Principle}(b)), which in turn perturb the optical readout due to the optomechanical coupling (Fig. \ref{fig:Principle}(c)) \cite{Han2016}. 
Previously, sensing of fluid density and speed of sound \cite{Bahl2013a,HyunKim2013}, fluid viscosity \cite{Han2014c}, and flowing particles \cite{Han2016, Suh2017} have been demonstrated using this platform. 
A recent report \cite{Suh2017} showed particle detection rate of exceeding 50,000 particles-per-second, without any binding, labeling, or reliance on random diffusion. It has also been shown \cite{Han2016,Suh2017a} that the size, position, density, and compressibility of the flowing particles, and the vibrational modeshape of the OMFR, all influence such measurements. 
The contributions of these individual parameters can be modeled using a Helmholtz equation  \cite{Han2016}, resulting in the a linearized model in the limit of small perturbations as follows:  
\begin{align}
\begin{split}
\frac{\Delta f}{f_1}= - \frac{\kappa_s - \kappa_{\ell}}{2\kappa_{\ell}}A -\frac{\rho_s - \rho_{\ell}}{2\rho_s}B, 
\label{eqn:Taylor}
\end{split}
\end{align}
where $A = \frac{\langle W_{sp} \rangle}{\langle W_{\ell p} \rangle}$, $B=\frac{\langle W_{sk}\rangle}{\langle W_{\ell p}\rangle}$. $\Delta f = f_2 -f_1$ is the frequency difference between the perturbed situation i.e. with a particle present (denoted by subscript ``2") and the unperturbed situation (denoted by subscript ``1").
The subscript ``s" denotes the particle such that $\kappa_s$ ($V_s$) represents the particle compressibility (volume). The subscript ``$\ell$" denotes the liquid core such that $\kappa_{\ell}$ ($\rho_{\ell}$) is the compressibility (density) of the fluid.
For the OMFR mechanical mode under consideration $W_{sp}$ ($W_{sk}$) is the acoustic potential (kinetic) energy in the fluid, but evaluated only over the volume displaced by the particle. 
$W_{{\ell}p}$ ($W_{{\ell}k}$) is the acoustic potential (kinetic) energy in the entire liquid core %\comment{... excluding the particle?} 
including the particle. 
The operation $\langle \boldsymbol{\cdot} \rangle$ denotes the time average over the period of oscillation $T$ as $\frac{1}{T} \int_{t}^{t+T} \, \boldsymbol{\cdot} \, dt$.
Therefore, $A$ and $B$ represent  sensitivity factors of the OMFR to the added particle and are both spatially dependent.

The above model is derived using the Helmholtz equation with the consideration of the fluid portion of the resonator cavity alone, which does not incorporate the acoustic pressure  modification by the particle, i.e. the scattering effect.
To more accurately predict the frequency shift, we must incorporate both the pressure field modification by the particle and the  energy change of the resonator shell into our model. 
We thus derive a new perturbation model below, in which we leverage energy conservation, including all energy components of the resonant motion. Using this model, we will be able to quantify how the change of each parameter (i.e. particle size, material properties, acoustic field etc.) contributes to the resonant frequency change. Moreover, we will be able to explicitly model the contribution from the particle  properties  and from the acoustic scattering effect.

\begin{figure}[htbp]
	\centering
	\includegraphics[width=0.38\textwidth, clip=true, trim=0.75in 7.45in 3.in 0in]{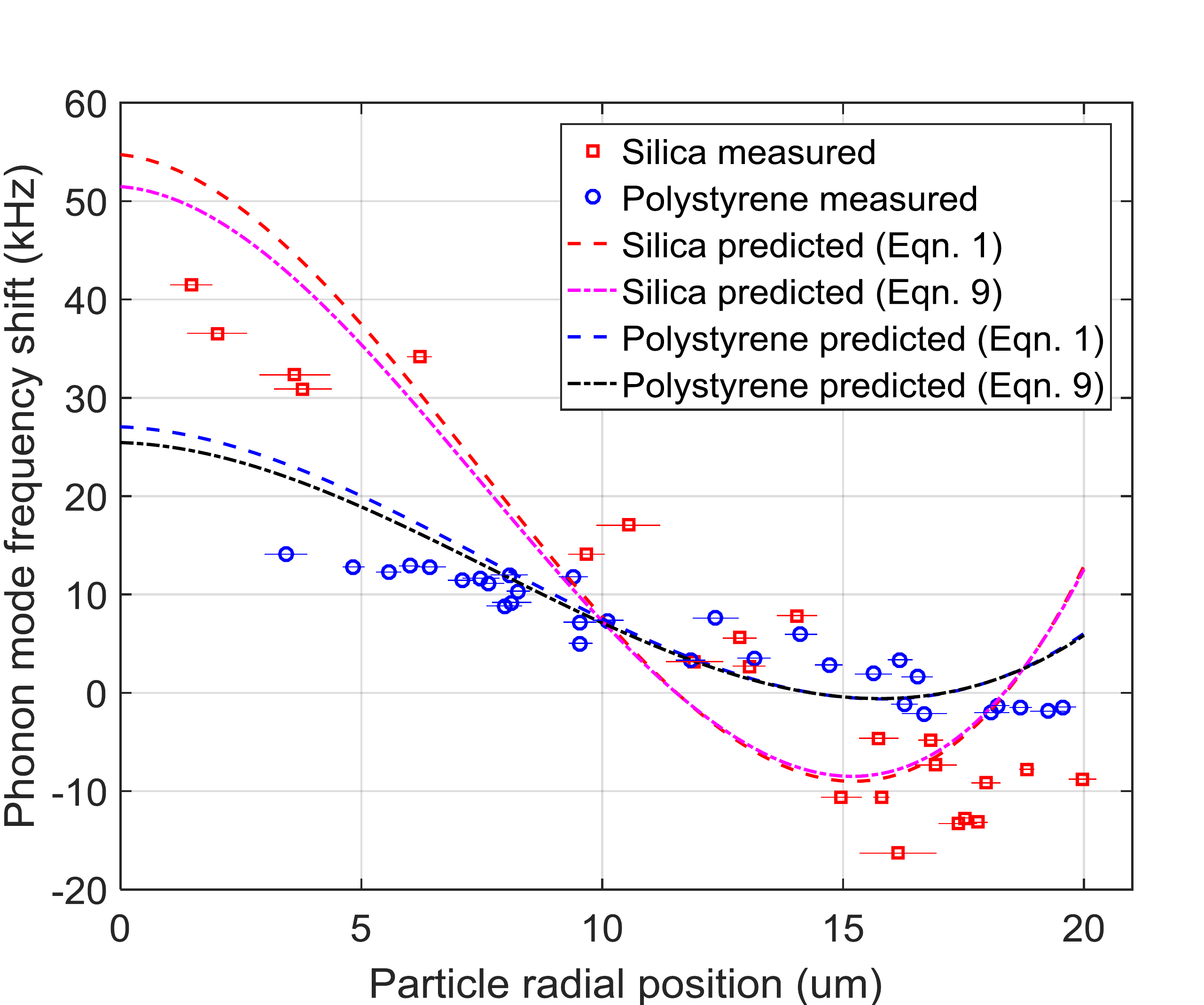}
	\caption{{\bf  } 
		Experimentally measured frequency shifts of the phonon mode, using 6 \um polystyrene and silica particles as perturbers. Lines represent analytical predictions.   The particle position is subject to both the fitting error (shown here by the error bar) and a roughly 2 \um error in determining the central axis of the OMFR \cite{Han2016}.  } 
	\label{fig:KappaResult}
\end{figure}

%% INTENTIONALLY BROKE PARAGRAPH HERE
For a harmonic oscillator, the average kinetic energy per oscillation cycle must equal to the average potential energy per oscillation cycle. For the OMFR, 
when there is no particle in the resonator, this relation can be expressed as:
\begin{align}
\langle W_{wp} \rangle + \langle W_{{\ell}p} \rangle =\langle W_{wk} \rangle + \langle W_{{\ell}k} \rangle, 
\label{eqn:S1}
\end{align}
where $W_{wp}$ ($W_{wk}$) is the elastic strain (kinetic) energy {associated with} the resonator shell or wall ``w''.
We can then express this unperturbed situation as follows:
\begin{multline}
	\langle W_{wp1} \rangle + \Big \langle \int_{V_{\ell}}\frac{1}{2}\kappa _{\ell} {P_1^2} dV \Big \rangle =  \\
	\Big \langle \int_{V_{w}}\frac{1}{2} \rho_{w}\Omega_1^2 |\vec{U}_1|^2 dV \Big \rangle + \Big \langle 	\int_{V_{\ell}} \frac{1}{2} \frac{|\nabla P_1|^2}{\rho_{\ell} \Omega_1^2} dV \Big \rangle,
	\label{eqn:eqn3}   
\end{multline}
where  $\rho_w$ ($V_w$) is the shell material density (shell volume), $V_{\ell}$ is the {resonator's} fluid core volume,  and $\Omega_i$ is the frequency of the vibrational mode. 
Again, the subscript ``1'' denotes the unperturbed case.
By assuming the oscillation to be time harmonic, i.e. the elastic displacement of the shell {is} $\vec{U_i}(r,t) = \vec{u_i}(r)\cos(\Omega_i t) $  and the pressure field $P_i(r,t) = p_i(r)\cos(\Omega_i t)$, we can further simplify Eqn. \ref{eqn:eqn3} {to}: 
\begin{multline}
	\langle W_{wp1} \rangle +\frac{\kappa_{\ell}}{4}\int_{V_{\ell}}p_1^2 dV=\\
	\frac{1}{4}\rho_{w}\Omega_1^2 \int_{V_{w}}  |\vec{u}_1|^2 dV+ \frac{1}{4} \frac{1}{\rho_{\ell} \Omega_1^2}\int_{V_{\ell}} |\nabla p_1|^2 dV.
	\label{eqn:caseI} 
\end{multline}
{We now place} a particle of volume $V_s$ inside the resonator {liquid volume} at some {fixed} location 
{and} rewrite this energy balance as:
\begin{multline}
	\langle W_{wp2} \rangle +\frac{\kappa_{\ell}}{4}\int_{V_{\ell}-V_s}p_2^2 dV 	+\frac{\kappa_s}{4}\int_{V_s}p_2^2 dV=\\	\frac{1}{4}\rho_{w}\Omega_2^2 \int_{V_w}  |\vec{u}_2|^2 dV	+ \frac{1}{4} \frac{1}{\rho_{\ell} \Omega_2^2}\int_{V_{\ell}-V_s} |\nabla p_2|^2 dV+ \\
	\frac{1}{4} \frac{1}{\rho_s \Omega_2^2}\int_{V_s} |\nabla p_2|^2 dV. 
	\label{eqn:caseII} 
\end{multline}
For  added simplicity, we further assume that the elastic displacement field of the shell does not change when the particle is added, i.e. $\vec{u}_1=\vec{u}_2=\vec{u}$, and thus $ \langle  W_{wp1} \rangle=\langle W_{wp2}\rangle$. 
We can then subtract Eqn. \ref{eqn:caseII} from Eqn. \ref{eqn:caseI} and obtain:
\begin{multline}
	%\begin{multline}
	%\begin{align}
	%\begin{split}
	\frac{\kappa_{\ell}}{4}(T_1-T_2) +  \frac{1}{4}(\kappa_{\ell} t_1 - \kappa_s t_2)\\
	 - \frac{1}{4 \rho_{\ell} \Omega_1^2}(G_1-G_2) - \frac{1}{4} \left(\frac{g_1}{\rho_{\ell} \Omega_1^2} - \frac{g_2}{\rho_s \Omega_2^2} \right) =\\
	\frac{F}{4}(\Omega_1^2 - \Omega_2^2) + \frac{G_2}{4 \rho_{\ell}} \left(\frac{1}{\Omega_1^2} -\frac{1}{ \Omega_2^2} \right), 
	%\end{split}
	%\end{align}
	%\end{multline}
\end{multline}
where we have {introduced} the following notations: $T = \int_{V_{\ell}-V_s} p^2 dV$, $t = \int_{V_s} p^2 dV$, $G = \int_{V_{\ell}-V_s} |\nabla p|^2 dV$, $g = \int_{V_s} |\nabla p|^2 dV$, and $F = \rho_{w} \int_{V_w}  |\vec{u}|^2 dV$. Since the frequency perturbation is {assumed to be} small compared with the resonance frequency, we have: 
$\frac{g_1}{\rho_{\ell} \Omega_1^2} - \frac{g_2}{\rho_s \Omega_2^2} \approx \frac{1}{\Omega_1^2} \left(\frac{g_1}{\rho_{\ell} } - \frac{g_2}{\rho_s } \right)$, and $\Omega_1^2 - \Omega_2^2 \approx - 2 \Omega_1 \Delta \Omega$, where $\Delta \Omega = \Omega_2 - \Omega_1$. We then obtain the fractional frequency perturbation {of} the mechanical mode as:
\begin{multline}
\frac{\Delta \Omega}{\Omega_1}= 
	\frac{1}{  - \frac{\Omega_1^2 F}{2}   + \frac{G_2} {2 \rho_{\ell} \Omega_1^2}  }  
	\biggl(  \frac{\kappa_{\ell}}{4}(T_1-T_2) +  \frac{1}{4}(\kappa_{\ell} t_1 - \kappa_s t_2) \\ 
	- \frac{1}{4 \rho_{\ell} \Omega_1^2}(G_1-G_2) - \frac{1}{4 \Omega_1^2}(\frac{g_1}{\rho_{\ell} } - \frac{g_2}{\rho_s }) \biggr).
\label{eqn:fullenergy} 
\end{multline}
The terms $T_1 - T_2$ and $G_1 - G_2$ {represent the liquid volume excluding the particle, and} are {thus} non-zero due to the scattering effect. {The effect is subtle but still observable in}  the FEM simulation in Fig. \ref{fig:Principle}(a) and (b). {In theory,} by simulating both the perturbed and unperturbed situations,  we can use Eqn. \ref{eqn:fullenergy} to calculate the frequency perturbation {quite precisely}.
However, the computational model has to be accurate enough to get a good estimation of all the difference terms in the numerator of Eqn. \ref{eqn:fullenergy}, which is challenging and computationally expensive. 
Nonetheless, if we can theoretically calculate the scattering field in the future, we should be able to get a much better prediction.  
Presently, we can further simplify Eqn. \ref{eqn:fullenergy} by neglecting the scattering effects such that that $p_1 = p_2 = p$, then $T_1 = T_2 $,  $G_1 =G_2$, $t_1 = t_2= t$, and $g_1 = g_2= g$. 
We then obtain:
\begin{align}
	\begin{split}
	\frac{\Delta \Omega}{\Omega_1}= 
	\frac{    \frac{t}{4}(\kappa_{\ell}  - \kappa_s ) - \frac{g}{4 \Omega_1^2}(\frac{1}{\rho_{\ell} } - \frac{1}{\rho_s }) }  {  - \frac{\Omega_1^2 F}{2}   + \frac{G}{2 \rho_{\ell} \Omega_1^2}}. 
	\end{split}
	\label{eqn:fullenergy1} 
\end{align}
Finally, by defining $\Omega_2 = 2 \pi f_2$ and $\Omega_1 = 2 \pi f_1$, and treating $G = \int_{V_{\ell}-V_s} |\nabla p|^2 dV \approx \int_{V_{\ell}} |\nabla p|^2 dV$ for the denominator since the particle is much smaller compared to the mode volume, we obtain the perturbed resonance frequency
\begin{align}
	\begin{split}
	\frac{\Delta f}{f_1}= -\frac{\kappa_s-\kappa_{\ell}}{2 \kappa_{\ell}} C
	-\frac{\rho_s-\rho_{\ell}}{2 \rho_s} D, 
	\end{split}
	\label{eqn:fullenergy2} 
\end{align}
where $C= \frac{ \langle W_{sp} \rangle}{ \langle W_{{\ell}k} \rangle - \langle W_{wk} \rangle}$, $D=\frac{ \langle W_{sk} \rangle}{ \langle W_{{\ell}k} \rangle - \langle W_{wk}\rangle}$. 
As we can see, the frequency perturbation prediction equations derived from the energy balance (Eqn. \ref{eqn:fullenergy2}) and the Helmholtz equation (Eqn. \ref{eqn:Taylor}) have very similar form. Essentially, the frequency perturbation occurs due to the modification of system potential energy through compressibility contrast and the modification of system kinetic  energy through density contrast. 
The energy term in the denominator is the only difference given by the two methods.  The influence this difference {makes} in the frequency perturbation predictions is small (for resonators with thin shell) as shown later in Fig. \ref{fig:KappaResult}.

{In order to verify the theoretical model, we performed a series of experiments}  {on} 6  \um polystyrene particles and  6 \um silica particles (Corpuscular 100235-10). The OMFR {that we} used has approximate maximum outer diameter of  55 \um  and estimated wall thickness of 4.5 $\mu$m, and supports a 34 MHz breathing vibrational mode. The experimental setup has been described in detail previously in \cite{Han2016}  (Fig. \ref{fig:Principle} (c)).  
The material  properties of the particles \cite{Lide1991}  used are summarized in Table \ref{tab:kappa}, where $c_P$ ($c_S$) is the P-wave (S-wave) acoustic velocity. 
{Measurements of} the frequency perturbations {generated} by {each of} these particles are shown in Fig. ~\ref{fig:KappaResult}, as a function of particle radial location relative to the OMFR center axis.

\begin{table}[htbp]
	\centering
	\caption{Material properties \cite{Lide1991}}  
	\begin{tabular}{ccccc}
		%\toprule
		\hline
		{Material} & {$\rho$ (kg/m$^3$)} & {$c_P$ (m/s)} & {$c_S$ (m/s)} & {$\kappa$ (Pa$^{-1}$)} \\
		\hline
		
		{Water} & 1000 & 1482 &  NA & \num{4.55e-10}\\
		\hline		
		{Silica} & 2200 & 5968 & 3764 & \num{2.72e-11} \\
		\hline
		{Polystyrene} & 1060 & 2350 & 1120 & \num{2.45e-10} \\
		\hline		
		%\bottomrule
	\end{tabular}%
	\label{tab:kappa}%
\end{table}%

\begin{figure}[htbp]
	\centering
	\includegraphics[width=0.43\textwidth, clip=true, trim=0in 9.75in 5.5in 0.1in]{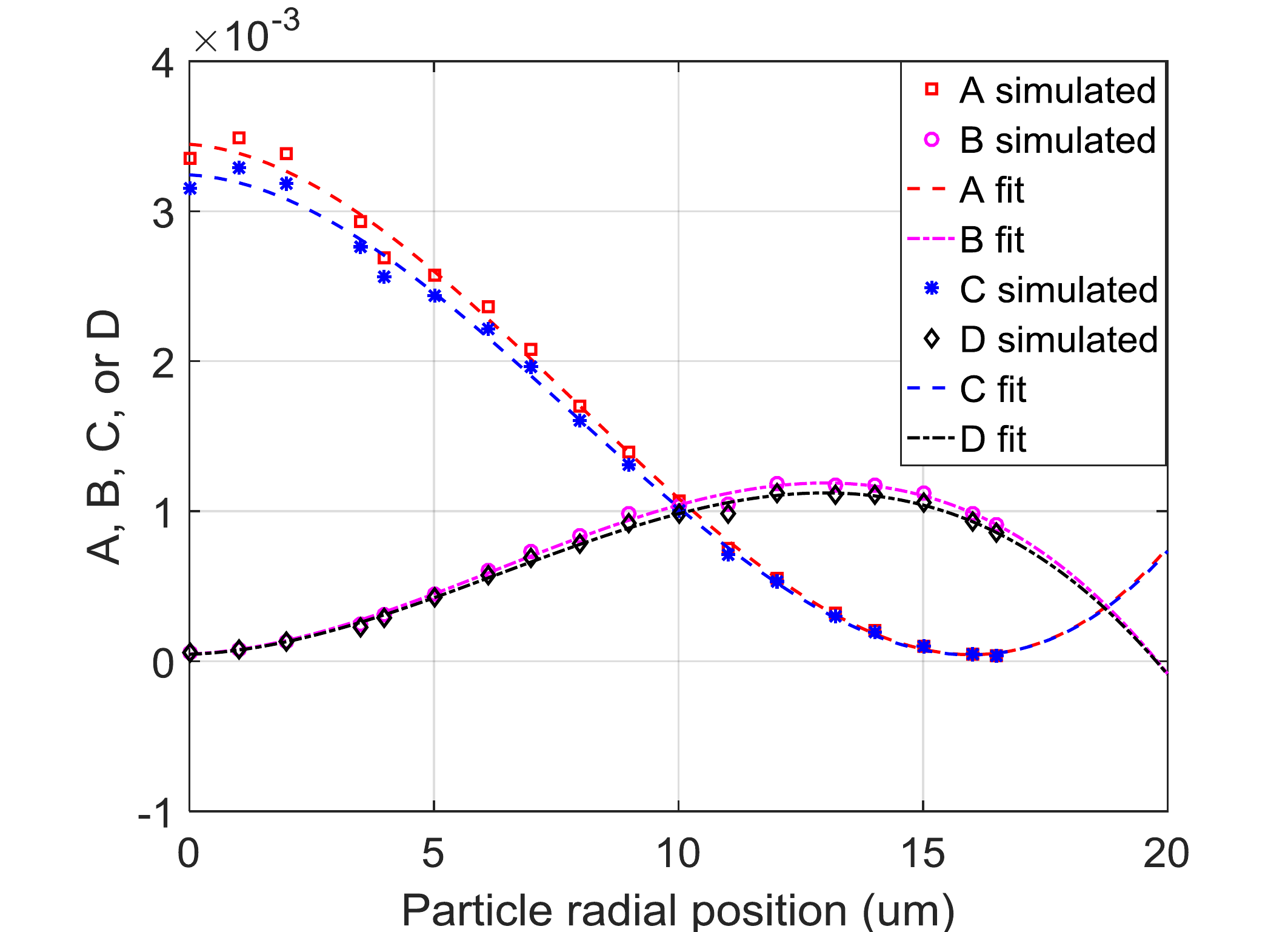}
	\caption{{\bf  } 
		Sensitivity variables $A$, $B$, $C$, and $D$ (Eqns.~\ref{eqn:Taylor} and \ref{eqn:fullenergy2})  obtained from simulation, along with polynomial curve fits.} 
	\label{fig:ABCD}
\end{figure}

To compare the results between the Helmholtz method and the energy balance, we first performed a {finite-element} simulation  {of the unloaded resonator using the method described} in \cite{Zhu2014}. 
The simulation results {are used to estimate the $A$, $B$,  $C$, and $D$ parameters as a function of particle location} (Fig. \ref{fig:ABCD}). 
A third-order-polynomial  curve fit is {generated for each parameter}. 
These  {fits}  are then used to predict the frequency perturbation using Eqn.~\ref{eqn:fullenergy2} and Eqn.~\ref{eqn:Taylor}, which are also plotted in Fig. \ref{fig:KappaResult}. 
While the predictions from {either model} can be used to qualitatively describe  the experimental results, there is appreciable  deviation between the predictions and the measurements -- the prediction  overestimates frequency shifts when the particles are close to the center axis of the OMFR and underestimates frequency shifts when particles are away from the center axis.

\begin{figure}[htbp]
	\centering
	\includegraphics[width=0.43\textwidth, clip=true, trim=0.25in 3.2in 4.5in 0in]{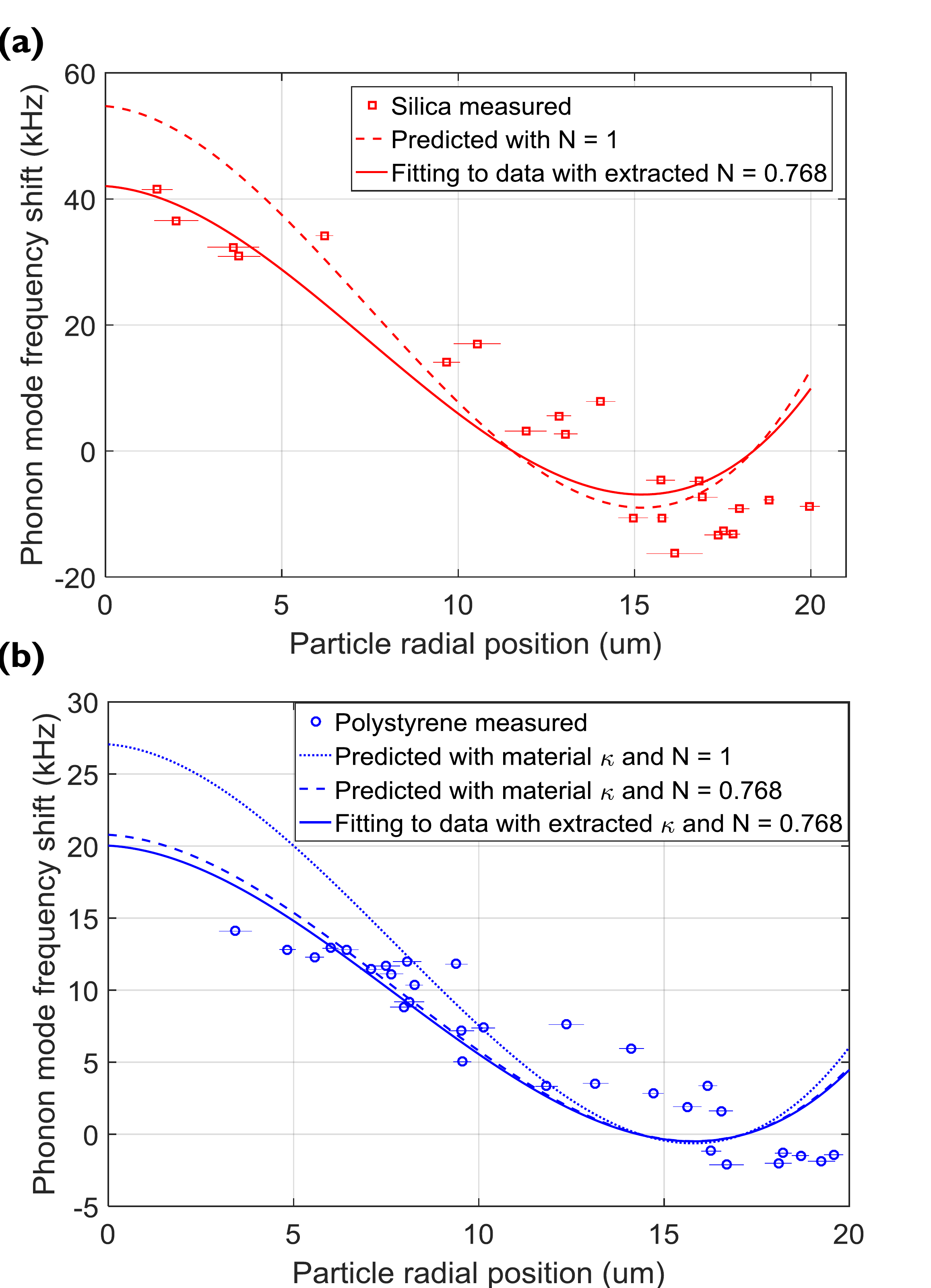}
	\caption{{\bf  } 
		(a)	Least-square curve fitting for Eqn.~\ref{eqn:N} is applied to the silica data to obtain $N = $ 0.768. The material properties are listed in Table \ref{tab:kappa}. The prediction with $N = $ 1 is given as a comparison. (b) The obtained $N$  is now used for least-squares curve fitting  to the polystyrene data with the compressibility of polystyrene as the unknown variable. Fitting to the experimental data extracts the compressibility of polystyrene as \num{2.53e-10} Pa$^{-1}$,  which is only 3\% different  from the data in Table \ref{tab:kappa}. The predictions with compressibility from Table \ref{tab:kappa}, with and without  scaling factor $N$, are given as comparisons.  } 
	\label{fig:KappaMethodI}
\end{figure}

We note that the mismatch between the experimental data and predictions has similar feature for both  silica and polystyrene particles.
We thus  propose to utilize one of the particles for calibrating  the system{,} and that a single empirical scaling parameter $N$ can be  introduced into  Eqn. \ref{eqn:Taylor} such that:
\begin{align}
	\Delta f= Nf_1 \left(-\frac{\rho_s - \rho_{\ell}}{2\rho_s}B - \frac{\kappa_s - \kappa_{\ell}}{2\kappa_{\ell}}A \right)
	\label{eqn:N}
\end{align}
In this study, we {have selected to use} the 6 \um silica particle{s} for calibration to obtain $N$.  We {first} perform a least-square curve fit using the silica test data with $N$ as an unknown {parameter} and all the other parameters as known (Fig. \ref{fig:KappaMethodI}(a)). Here, we use the material properties from Table \ref{tab:kappa} and $N = $ 0.768  $\pm$ 0.171 is obtained. 
We then use this empirically extracted $N$ as a known parameter and apply the least-square curve fit for Eqn. \ref{eqn:N} again to a set of polystyrene experimental data with the compressibility of polystyrene as the unknown {parameter} (Fig. \ref{fig:KappaMethodI}(b)). 
This fit allows us to estimate the compressibility of polystyrene as (2.53 $\pm$ 0.26)$\times$\num{e-10}  Pa$^{-1}$.  The error compared with data provided in Table \ref{tab:kappa} is  3$\%$. 
As a comparison, if we perform a curve fit for Eqn.~\ref{eqn:N} again but with $N=$ 1, the compressibility extracted is (2.97 $\pm$ 0.20)$\times$\num{e-10}  Pa$^{-1}$, which is  22$\%$ error  compared with data provided in Table \ref{tab:kappa}. We thus conclude that our single-parameter empirical method has improved the compressibility prediction  significantly.

We note that the above method cannot provide accurate sensitivity prediction when the {compressibility} % $\kappa$ 
of the particle is much smaller than that of the ambient fluid. In other words, such particles are extremely rigid with respect to their environment. This can be seen from Eqn. \ref{eqn:N} {by considering the factor} $\frac{\kappa_s - \kappa_{\ell}}{2\kappa_{\ell}} = \frac{\kappa_s}{2\kappa_{\ell}}- \frac{1}{2}$, such that for $\kappa_s << \kappa_{\ell}$ a small error in the compressibility contrast due to fitting results in a large relative error in the estimation of $\kappa_s$.  
To illustrate this limitation, we now perform a curve fitting with Eqn. \ref{eqn:N} to the  silica experimental data  (which has much lower compressibility than water) to extract the compressibility of silica. However, this time we employ the compressibility of silica as the unknown variable and use $N=$ 0.768 as a known parameter. Since the scaling factor $N$ was found using the density and compressibility of silica directly, we can therefore minimize the mismatch between experimental data and theoretical prediction, and find the best possible compressibility estimation using current approach.  
The compressibility for silica extracted by curve fitting is (1.75 $\pm$ 9.00)$\times$\num{e-11}  Pa$^{-1}$. The impractically large error bar indicates that we simply cannot trust this extracted value.

The technique that we have demonstrated in this work permits the measurement of compressibility for single particles in fluid media, with high throughput, without contact or labeling. The approach is particularly powerful for measurements on soft particles, or biological particles, that have relatively high compressibilities and for which analysis of large population statistics is critical.

\begin{acknowledgments}
Funding for this research was provided by the US National Science Foundation (grants ECCS-1509391 and ECCS-1408539).
\end{acknowledgments}

\bibliography{phdthesis1}

\end{document}